\documentclass[dvips]{article}

\usepackage{icrc2011}
\usepackage{unites2e}
\usepackage{graphicx}
\usepackage{url}

\title{VERITAS: Status and Highlights}

\newcommand{\etal}{\MakeLowercase{\textit{et al. }}} 
\shorttitle{ J. Holder \etal VERITAS: Status and Highlights}
\authors{
J.~Holder$^{1}$ for the VERITAS Collaboration$^{2}$:   
E.~Aliu,     
T.~Arlen,    
T.~Aune,     
M.~Beilicke,  
W.~Benbow,    
M.~B{\"o}ttcher, 
A.~Bouvier,    
J.~H.~Buckley, 
V.~Bugaev,    
K.~Byrum,    
A.~Cannon,     
A.~Cesarini,    
J.~L.~Christiansen,
L.~Ciupik,    
E.~Collins-Hughes,
M.~P.~Connolly,
W.~Cui,     
R.~Dickherber,
C.~Duke,    
V.~V.~Dwarkadas,
M.~Errando,   
A.~Falcone,   
J.~P.~Finley, 
G.~Finnegan, 
L.~Fortson,  
A.~Furniss,  
N.~Galante,  
D.~Gall,     
K.~Gibbs,    
G.~H.~Gillanders, 
S.~Godambe,  
S.~Griffin,  
J.~Grube,    
R.~Guenette, 
G.~Gyuk,     
D.~Hanna,    
J.~Holder,   
H.~Huan,      
G.~Hughes,    
C.~M.~Hui,    
T.~B.~Humensky,
A.~Imran,     
P.~Kaaret,    
N.~Karlsson,  
M.~Kertzman,  
Y.~Khassen,   
D.~Kieda,     
H.~Krawczynski, 
F.~Krennrich,  
M.~J.~Lang,    
M.~Lyutikov,   
A.~S~Madhavan, 
G.~Maier,      
P.~Majumdar,   
S.~McArthur,   
A.~McCann,     
M.~McCutcheon, 
P.~Moriarty,   
R.~Mukherjee,  
P.~D~Nu\~{n}ez, 
R.~A.~Ong,  
M.~Orr,     
A.~N.~Otte,  
N.~Park,     
J.~S.~Perkins,
A.~Pichel,   
F.~Pizlo,    
M.~Pohl,     
H.~Prokoph,  
J.~Quinn,    
K.~Ragan,     
L.~C.~Reyes,   
P.~T.~Reynolds,
A.~C.~Rovero,  
J.~Ruppel,     
A.~C.~Sadun,     
D.~B.~Saxon,      
M.~Schroedter,      
G.~H.~Sembroski,     
G.~D.~\c{S}ent\"{u}rk,
A.~W.~Smith,   
D.~Staszak,    
G.~Te\v{s}i\'{c},
M.~Theiling,   
S.~Thibadeau,  
K.~Tsurusaki,  
J.~Tyler,      
A.~Varlotta,   
V.~V.~Vassiliev,
S.~Vincent,   
M.~Vivier,    
S.~P.~Wakely, 
J.~E.~Ward,   
T.~C.~Weekes, 
A.~Weinstein, 
T.~Weisgarber,
D.~A.~Williams,
B.~Zitzer     
}

\afiliations{

$^1$ Department of Physics and Astronomy and the Bartol Research Institute, University of Delaware, Newark, DE 19711.\\
$^2$see http://veritas.sao.arizona.edu/conferences/authors?icrc2011 for a full list of affiliations\\
}
\email{jholder@physics.udel.edu}

\abstract{The VERITAS telescope array has been operating smoothly
since 2007, and has detected gamma-ray emission above 100 GeV from 40
astrophysical sources. These include blazars, pulsar wind nebulae,
supernova remnants, gamma-ray binary systems, a starburst galaxy, a
radio galaxy, the Crab pulsar, and gamma-ray sources whose origin
remains unidentified. In 2009, the array was reconfigured, greatly
improving the sensitivity. We summarize the current status of the
observatory, describe some of the scientific highlights since 2009,
and outline plans for the future.}


\begin{document}
\maketitle

 \begin{figure*}[!t]
  \centering
  \includegraphics[width=5.89in]{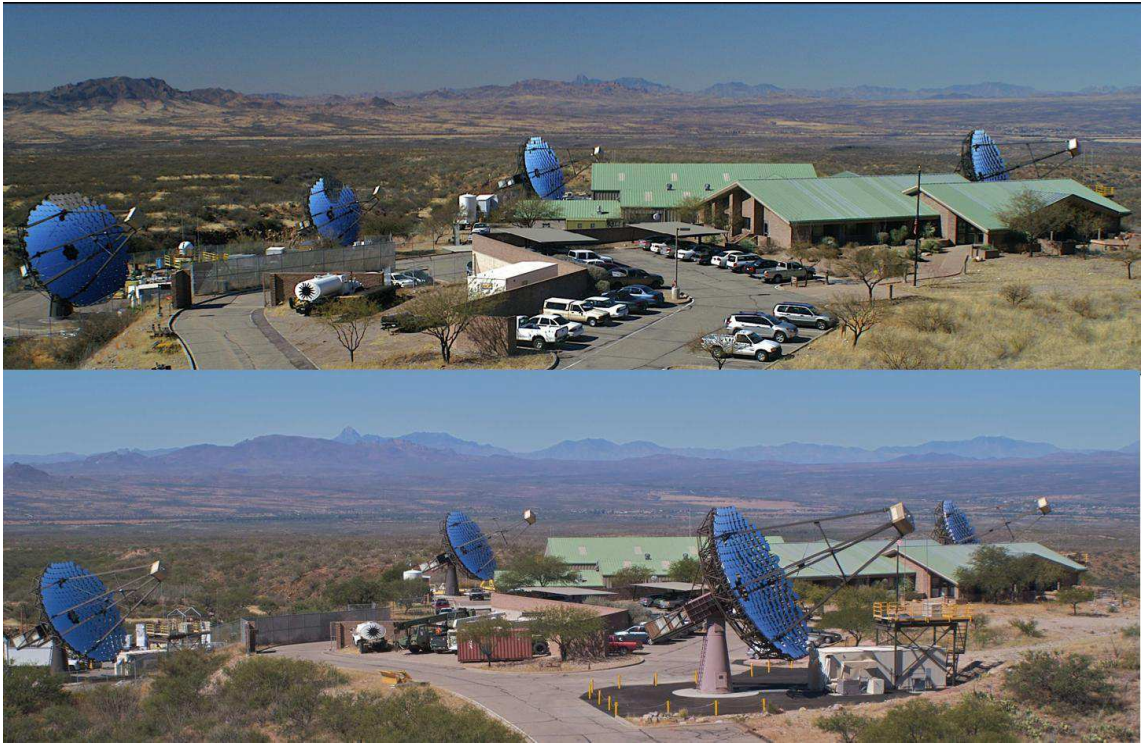}
  \caption{The VERITAS array of four gamma-ray telescopes in its
    initial ({\bf Top}) and present ({\bf Bottom}) configuration.  }
  \label{VERITAS}
 \end{figure*}

\section{Overview: VERITAS}

VERITAS (the \textit{Very Energetic Radiation Imaging Telescope Array
System}) is an array of four atmospheric Cherenkov telescopes located
near Tucson in southern Arizona ($31^{\circ}40'$N $110^{\circ}57'$W,
$1268\U{m}$ a.s.l) \cite{weekes02}. It is used to study astrophysical sources of
gamma-ray emission in the $100\U{GeV}$-$30\U{TeV}$ range via the
imaging atmospheric Cherenkov technique. The array was
commissioned in 2007, and the source catalogue now contains 40
objects, including pulsar wind nebulae (PWN), supernova remnants
(SNR), active galaxies, binary systems, a starburst galaxy and a
pulsar, along with other objects whose nature remains unclear.

Each of the four telescopes consists of a $12\U{m}$ diameter segmented
reflector, at the focus of which is a 499-pixel photomultiplier tube
(PMT) camera covering a field of view of $3.5^{\circ}$
\cite{holder06}. A coincident Cherenkov signal in 2 out of 4
telescopes triggers a read-out of the PMT signals by custom-built
$500\U{MSPS}$ FADCs, at a typical rate of $\sim250\U{Hz}$. The
resulting images in each camera are parameterized by their second
moments, and these parameters are then used to discriminate gamma-ray
initiated air showers from those initiated by cosmic ray
particles. The recorded images are also used to reconstruct the energy
and arrival direction of the primary photon. The angular resolution
and energy resolution of the reconstruction is energy dependent,
reaching $\sim0.1^{\circ}$ and $\sim15\%$, respectively, for gamma-ray
primaries with an energy of $1\U{TeV}$.

The sensitivity of the array can be quantified by the observing time
required to detect a typical weak source. The sensitivity of VERITAS
has steadily improved over the lifetime of the array, due to
improvements in data analysis techniques, optical alignment,
calibration and, most significantly, following the relocation of the
original prototype telescope to a more favourable location in
2009. Currently, a source with a flux of 1\% of the Crab Nebula flux
and a spectrum similar to the Crab Nebula can be detected in
approximately $25\U{hours}$ of observations: roughly half of the time
required when the array was originally
commissioned. Figure~\ref{VERITAS} shows the array in both its initial
(top) and current (bottom) configurations. Typically, around
$1000\U{hours}$ of data are collected every year, $\sim20\%$ of which
are taken when the moon is visible.

\section{Extragalactic TeV Gamma-ray Astronomy}
 \begin{figure*}[!t]
  \centering
  \includegraphics[width=5.89in]{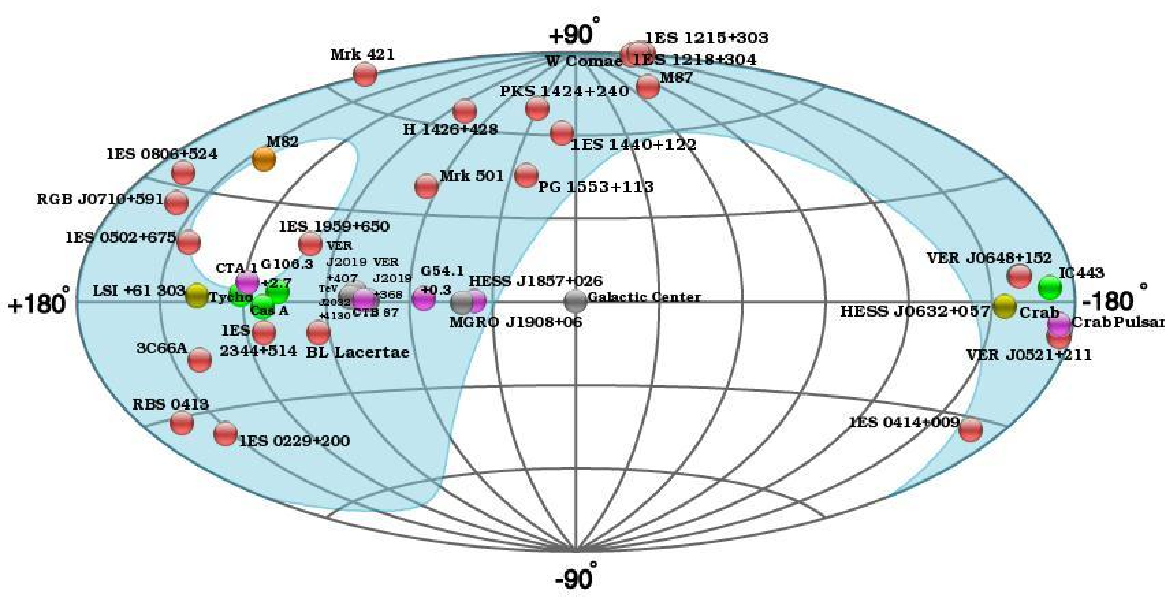}
  \caption{The VERITAS source map, in Galactic coordinates, as of July 2011.
    }
  \label{source_map}
 \end{figure*}

Approximately two-thirds of the VERITAS program of observations is
dedicated to the study of extragalactic gamma-ray sources and source
candidates. This focus has resulted in the detection by VERITAS of 21
blazars, one radio galaxy (M~87) and one starburst galaxy (M~82). The
extragalactic program is described in more detail elsewehere in these
proceedings~\cite{benbow, galante1}. Here, we highlight some of the
recent results and ongoing studies.

Table~\ref{extragalactic_sources} lists all of the extragalactic
sources detected, roughly half of which were first identified as
$>100\U{GeV}$ gamma-ray sources by VERITAS.

\begin{table}[!h]
\begin{center}
\begin{tabular}{|l|c|c|c|}
\hline
Source Name & Class & \textit{z}  \\
            &       &             \\
\hline
Mrk 421         & HBL    & 0.030  \\
Mrk 501         & HBL    & 0.034  \\
1ES 2344+514    & HBL    & 0.044  \\
1ES 1959+650    & HBL    & 0.047  \\
BL Lac          & LBL    & 0.069  \\
W Comae         & IBL    & 0.102  \\
RGB J0710+591   & HBL    & 0.125  \\
H 1426+428      & HBL    & 0.129  \\
1ES 0806+524    & HBL    & 0.138  \\
1ES 0229+200    & HBL    & 0.139  \\
1ES 1440+122    & IBL    & 0.162  \\
RX J0648.7+1516 & HBL    & 0.179  \\
1ES 1218+304    & HBL    & 0.182  \\
RBS 0413        & HBL    & 0.190  \\
1ES 0414+009    & HBL    & 0.287  \\
PG 1553+113     & HBL    & 0.43 $< z <$ 0.50  \\
1ES0502+675     & HBL    & ?  \\
3C 66A          & IBL    & ?  \\
B2 1215+30      & LBL    & ?  \\
PKS 1424+240    & I/HBL  & ?  \\
VER J0521+211   & HBL    & ?  \\
M~87             & FR I   & $16.7\U{Mpc}$  \\
M~82             & Starburst & $3.9\U{Mpc}$ \\
\hline
\end{tabular}
\caption{Extragalactic sources of TeV gamma-ray emission detected by VERITAS.}\label{extragalactic_sources}
\end{center}
\end{table}

\subsection{Blazars}

 \begin{figure}[!h]
  \vspace{0mm}
  \centering
  \includegraphics[width=3.2in,clip,trim = 1mm 1mm 0mm 1mm]{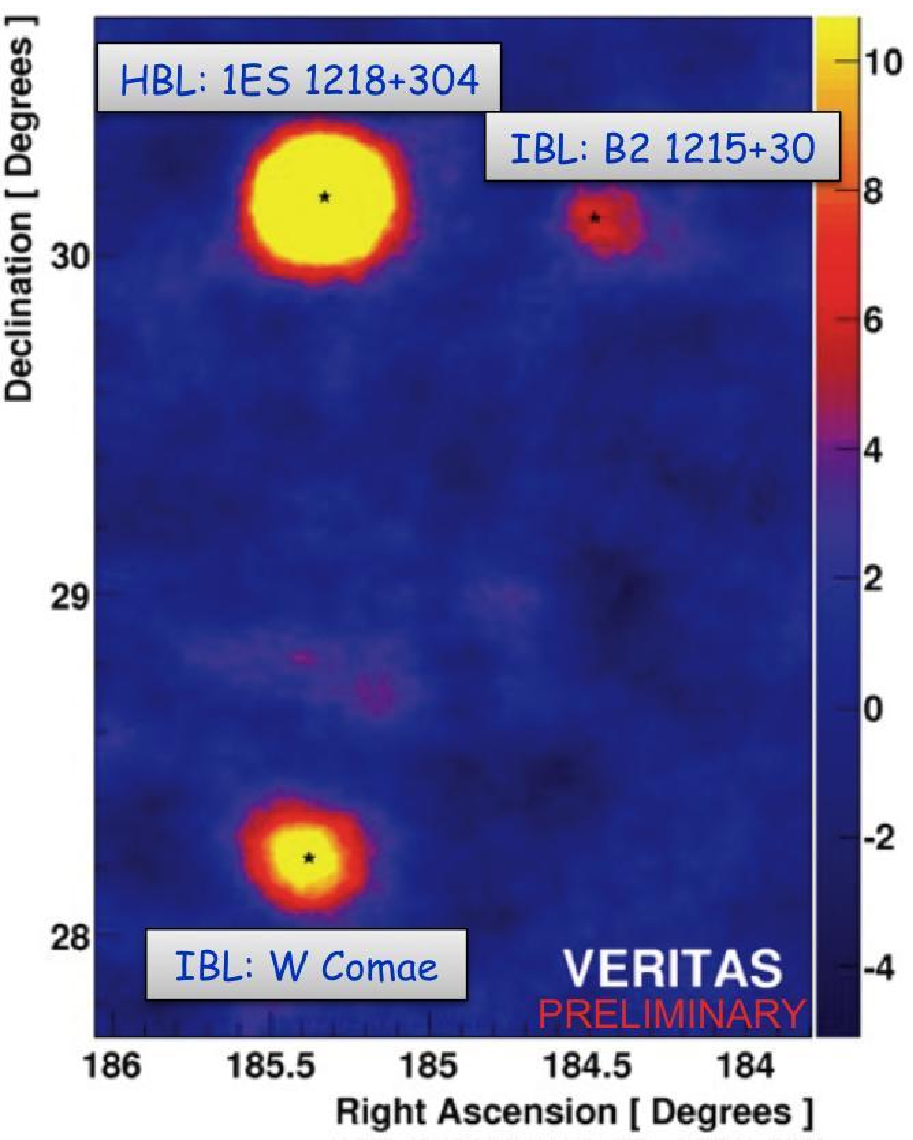}
  \caption{Significance map for a single field of view
  containing three blazars. Each source is point-like; the apparent
  size differences are due to saturation on the \textit{z}-scale}.
  \label{extragal_fig}
 \end{figure}

The most common extragalactic TeV sources are the blazars: Active
Galactic Nuclei (AGN) in which a relativistic jet is directed along
the line-of-sight to the observer. The first few TeV blazars to be
identified were all high-frequency-peaked BL Lac objects. The combined
results of H.E.S.S., MAGIC and VERITAS have dramatically expanded the
TeV blazar catalogue over the past few years, allowing population
studies to be carried out and adding new classes of
object. Originally, due mainly to limited sensitivity and biases in
target selection, only high-frequency peaked BL Lac objects were
detected at TeV energies. More recently, intermediate- and
low-frequency-peaked BL Lacs have been observed \cite{majumdar}, as
well as flat-spectrum radio quasars. The population is also
growing as the catalogue extends to ever larger distances. An
illustration of these developments is provided by
Figure~\ref{extragal_fig}, which shows VERITAS observations of three
blazars, one high-frequency and two intermediate-frequency BL Lac
objects, all contained within the $3.5^{\circ}$ field-of-view of the
array.

VERITAS observations of blazars have, in many cases, been guided and
enhanced by the results of observations at other wavelengths. In
particular, the \textit{Fermi}-LAT instrument provides a daily view of
the entire sky in the energy range just below that covered by VERITAS,
and the VERITAS blazar discovery program now focuses on objects
detected by \textit{Fermi}-LAT. Objects are typically selected from
the LAT catalog based either upon a power-law extrapolation of their
energy spectra, or upon the detection of clusters of high energy
($>10\U{GeV}$) photons by the LAT. We have also implemented analysis
pipelines that automatically process and analyze \textit{Fermi}-LAT data on a
daily basis, in order to identify flaring objects \cite{errando_orr}.

The Whipple 10m telescope has played an important role in the
blazar program, providing regular monitoring of bright, known, variable
blazars such as Mkn~421 and Mkn~501 \cite{pichel}. Summer 2011 marks
the end of operations for the Whipple 10m, which holds a remarkable
record as the longest operating atmospheric Cherenkov telescope
(since 1968), and as the instrument on which the Whipple Collaboration
developed the imaging technique used to detect the first TeV source,
the Crab Nebula \cite{weekes89}.

 \begin{figure*}[!t]
  \centering
  \includegraphics[width=5.in]{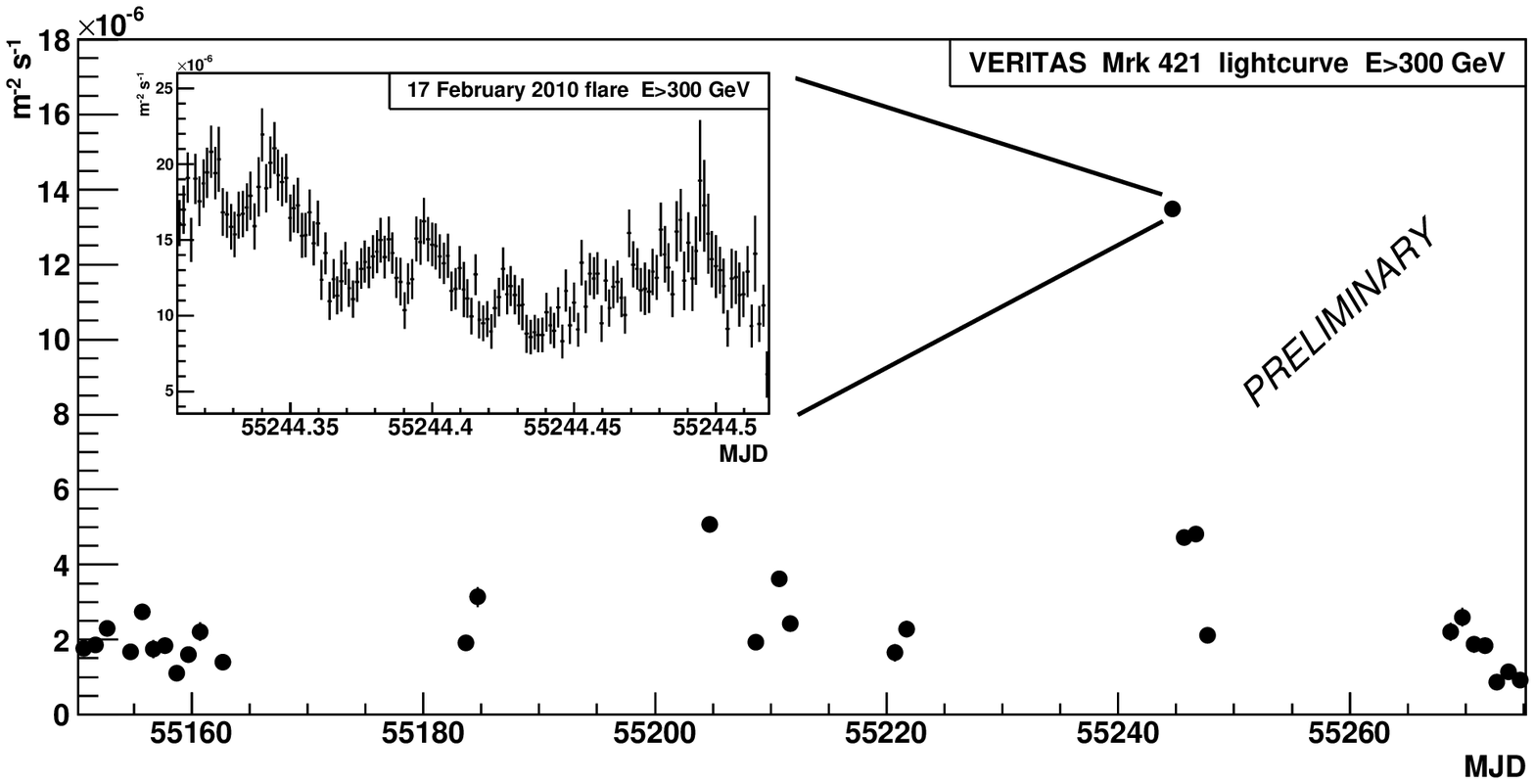}
  \includegraphics[width=5.in]{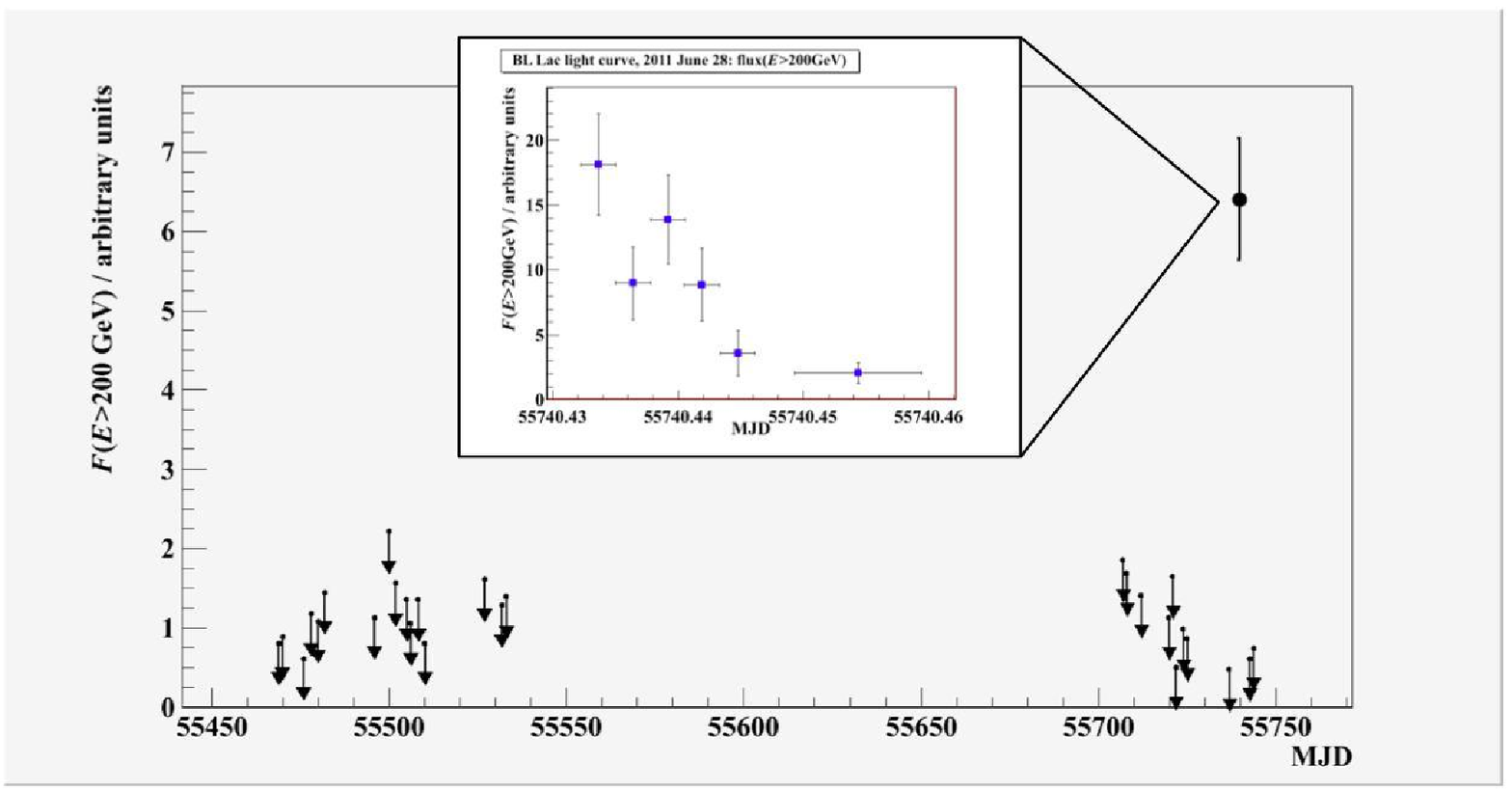}
  \caption{{\bf Top} Nightly lightcurve of Markarian 421 during the VERITAS 2009-2010 observing campaign, with a zoom of the flare on the night of February 17, 2010 \cite{galante2}. 
{\bf Bottom} Nightly lightcurve of BL Lacertae during the VERITAS 2010-2011 observing campaign, with a zoom of the flare on the night of June 28, 2011 \cite{bllac_atel}. 
    }
  \label{blazar_lightcurves}
 \end{figure*}

Highlights from the VERITAS blazar program presented at this
conference include the measurement of an extremely bright flare from
Mkn~421 in February 2010 \cite{galante2}. During $5\U{hours}$ of
VERITAS observations on February 17, gamma-ray emission from the
source reached a flux level of 8 times the steady flux from the Crab
Nebula, allowing precise measurement of the light curve with a time
resolution of two minutes (Figure~\ref{blazar_lightcurves}). We also
summarize observations of PG~1553+113 between May 2010 and May 2011
\cite{orr}. This source is one of the most distant TeV blazars known,
making it an important tool for studies of the extragalactic
background light. The VERITAS spectral measurements presented here are
used to place an upper limit on the source redshift of $z<0.5$ at 95\%
confidence. The identification of a new high-frequency-peaked blazar,
RBS~0413, with VERITAS and \textit{Fermi}-LAT was also presented
\cite{senturk}, providing a good example of the impact of the LAT upon
TeV target selection. Finally, observations of BL Lac (the eponymous
BL Lacertae object) were shown. BL Lac is the first blazar classified
as 'low-frequency-peaked' from which VERITAS has detected gamma-ray
emission. Following reports of activity from several observatories at
other wavelengths in May 2011, VERITAS commenced monitoring the source
and, on June 28, 2011, observed the tail-end of a flare during 40
minutes of observations, with further observations curtailed by the
rising sun. The flux decreased rapidly from a maximum of $\sim50\%$ of
the steady Crab Nebula flux, demonstrating variability on a
timescale of $\sim4\U{minutes}$
\cite{majumdar}(Figure~\ref{blazar_lightcurves}).

\subsection{Other Extragalactic Sources}

In addition to blazars, VERITAS has been used to observe various other
extragalactic gamma-ray source candidates, including radio galaxies,
galaxy clusters, starburst galaxies and globular clusters
\cite{galante1}. Highlights from this program include the study of
gamma-ray emission from M~87 and M~82, which we summarize here.

\subsubsection{M~87}

 \begin{figure*}[!t]
  \vspace{0mm}
  \centering
  \hspace{-5mm}
  \includegraphics[width=5in,clip,trim = 1mm 1mm 0mm 1mm]{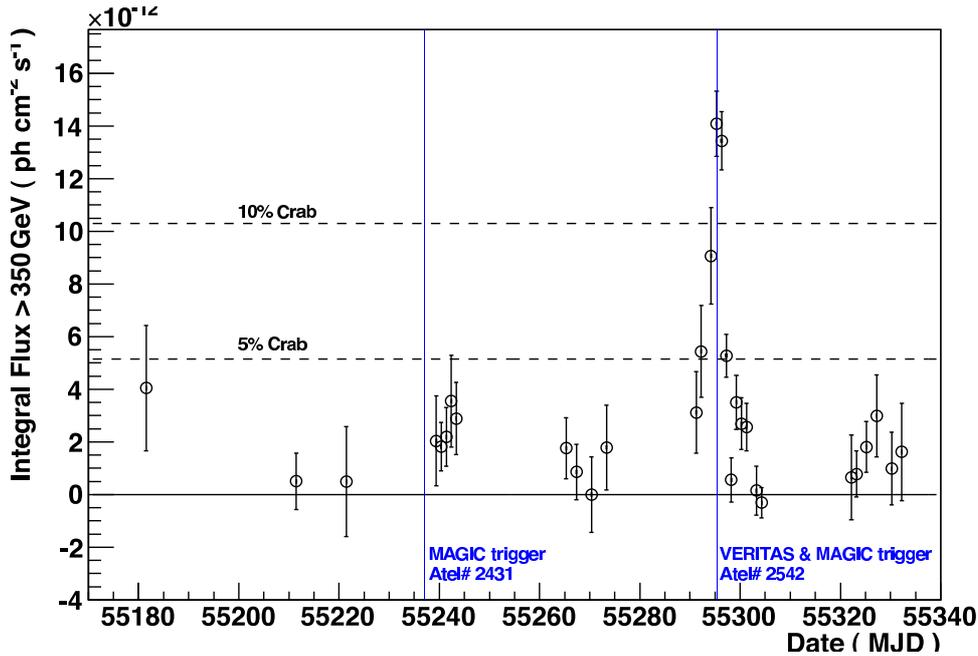}
  \caption{VERITAS lightcurve of M~87 in 2010, }.
  \label{M87}
 \end{figure*}

M~87 is the central giant radio galaxy of the Virgo cluster, located
at a distance of $16.7\U{Mpc}$ \cite{mei}. Its proximity, and the fact
that its jet is not aligned along the line-of-sight, allows detailed
study of the jet structure in radio, optical and X-ray
wavelengths. This provides the possibility of identifying which
structures in the jet are responsible for the TeV emission, and so
M~87 has been a prime target for gamma-ray observatories since the
initial report of gamma-ray emission by the HEGRA array in 2003
\cite{HEGRA_M87}. A concerted multiwavelength monitoring campaign
involving all of the TeV observatories, as well as Chandra, VLBA and a
number of optical observatories, has been ongoing for several years
(e.g. \cite{Science_M87}). Initial multiwavelength results indicated
that the gamma-ray emission may originate from a region close to the
core of the galaxy. In April 2010, the brightest flaring event to date
was observed, with a peak flux exceeding 10\% of the steady Crab
Nebula flux. The VERITAS lightcurve for 2010 is shown in
Figure~\ref{M87}, revealing flux variability during the flare on the
timescale of $\sim1\U{day}$.

\subsubsection{M~82}

M~82 is a bright galaxy located at a distance of approximately
$3.9\U{Mpc}$, with an active starburst region at its centre. The star
formation rate in this region is approximately 10 times that of the
Milky Way, with an estimated supernova rate of 0.1 to 0.3 per
year. High cosmic-ray and gas densities in the starburst region make
it a promising target for gamma-ray observations, with gamma-ray
emission expected to result from the interactions of hadronic cosmic
rays in the dense gas. A deep VERITAS exposure ($137\U{hours}$) in
2008-2009 resulted in a detection of gamma-ray emission from M~82 with
a flux of $(3.7 \pm 0.8_{stat} \pm 0.7_{syst}) \times 10^{-13}
\UU{cm}{-2} \UU{s}{-1}$ above the 700\U{GeV} energy threshold of the
analysis, consistent with the predictions of models based on the
acceleration and propagation of cosmic rays in the starburst core
\cite{Nature_M82}.

\subsubsection{Gamma-ray Bursts}
VERITAS has maintained a continuous program to search for very
high-energy emission associated with gamma-ray bursts (GRBs). The
motivation for such searches has been given added impetus in recent
years, due to the detection high energy emission from GRBs by
\textit{Fermi}-LAT with a delay of hundreds of seconds from the start
of the burst. VERITAS results at this conference focused on bursts
which were detected by both the \textit{Fermi} and \textit{Swift}
satellites \cite{aune}. Since VERITAS is a pointed instrument, with a
field of view of $3.5^{\circ}$, the telescopes must move into position
when a burst alert is received. The average time delay for this
repoint is $240\U{s}$, and in many cases significantly
shorter. Predictions based on the brightest bursts observed by the LAT
indicate that the potential for detecting TeV emission associated with
a GRB with VERITAS is promising, assuming no intrinsic spectral
cut-off of the high energy emission.

\subsubsection{Dark Matter Searches}
 \begin{figure*}[!t]
  \vspace{0mm}
  \centering
  \includegraphics[width=5in,clip,trim = 1mm 1mm 0mm 1mm]{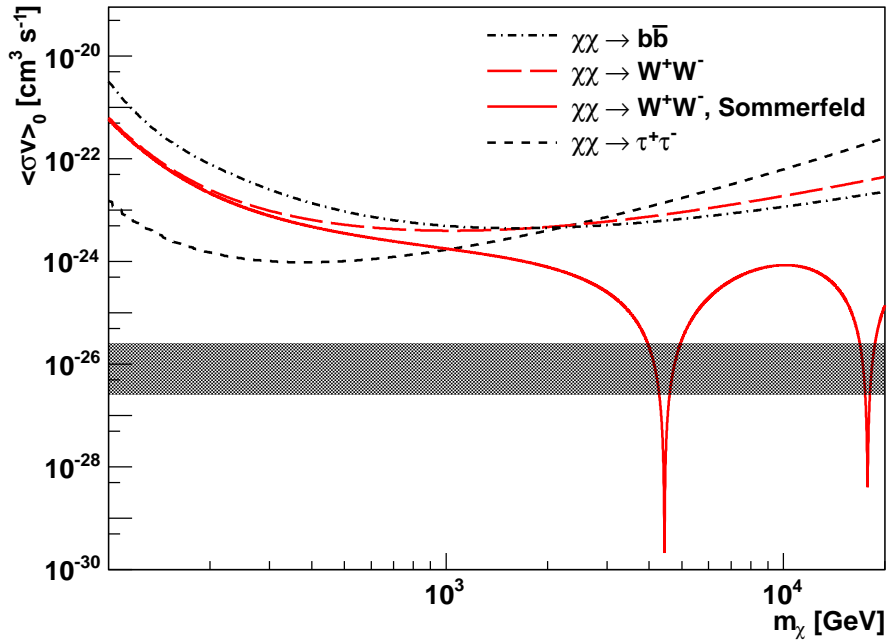}
  \caption{Upper limits at the 95\% confidence level on the
  velocity-weighted annihilation cross-section for different
  annihilation channels. The dark band represents the typical range of
  predictions.}
  \label{DM}
 \end{figure*}

Objects outside of our own galaxy provide some of the best
locations in which to search for the annihilation signatures of dark
matter particles. Dwarf spheroidal galaxies of the local group are
among the most promising of these, due to their proximity, their large
dark matter content and the absence of astrophysical background
sources (supernova remnants, pulsar wind nebulae, etc.). Studies of
the stellar kinematics of the dwarf spheroidal galaxy Segue I indicate
that is is the most dark matter dominated dwarf spheroidal galaxy
known. The results of a 48 hour VERITAS exposure on Segue I were
presented at this conference \cite{vivier}, along with upper limits on
the annihilation cross-section for various annihilation channels,
illustrated in Figure~\ref{DM}. The limits lie in the range
$10^{-22}-10^{-24} \UU{cm}{3} \UU{s}{-1}$, depending on the annihilation
channel and dark matter particle mass.

\section{Galactic TeV Gamma-ray Sources}

The VERITAS Galactic catalogue contains 17 sources, listed in
Table~\ref{galactic_sources}. VERITAS is located in the northern
hemisphere, and can therefore only view the outer Galaxy at high
elevation angles ($l<-30^{\circ}$ and $l>150^{\circ}$). The relative
scarcity of Galactic TeV sources in this region, compared to the inner
Galaxy, is balanced in part by the variety of source classes which
have been identified and studied. These include pulsars and their
nebulae, supernova remnants, gamma-ray binary systems, and a number of
sources whose identification remains unclear. Some of the recent
highlights from the Galactic program are summarized here. Additional
VERITAS results are presented elsewhere in these proceedings,
including observations of CTA 1 \cite{aliu}, LS~I+$61^{\circ}$303 and
1A~0535+262 \cite{maier2}, and VER~J2019+407 \cite{weinstein}.

\begin{table}[!h]
\begin{center}
\begin{tabular}{|l|c|c|c|}
\hline
Source Name & Class     \\
            & (likely) \\
\hline
Crab Nebula           & PWN      \\
Crab Pulsar           & Pulsar   \\
LS~I~+61$^{\circ}$303  & Binary   \\
IC~443                & SNR      \\
Cas~A                 & SNR      \\
G106.3+2.7            & SNR/PWN  \\
G54.1+0.3             & PWN      \\
HESS~J0632+057        & Binary   \\
CTA~1                 & PWN      \\
Tycho's SNR           & SNR      \\
HESS~J1857+026        & PWN?     \\
CTB~87                & PWN      \\
MGRO J1908+06         & PWN?     \\
TeV J2032+4130        & PWN?     \\
Galactic Centre       & UID      \\
VER~J2019+407         & UID      \\
Cyg OB1 TeV complex   & UID      \\
\hline
\end{tabular}
\caption{Galactic sources of TeV gamma-ray emission detected by
VERITAS. Some of the designations are not definitive; we list here the
most likely counterpart. The unidentified sources (UID) are presumed
Galactic due to their location in the Galactic plane or their
angular extent.}\label{galactic_sources}
\end{center}
\end{table}

\subsection{HESS J0632+057}

 \begin{figure*}[!t]
  \vspace{0mm}
  \centering
  \includegraphics[width=5in]{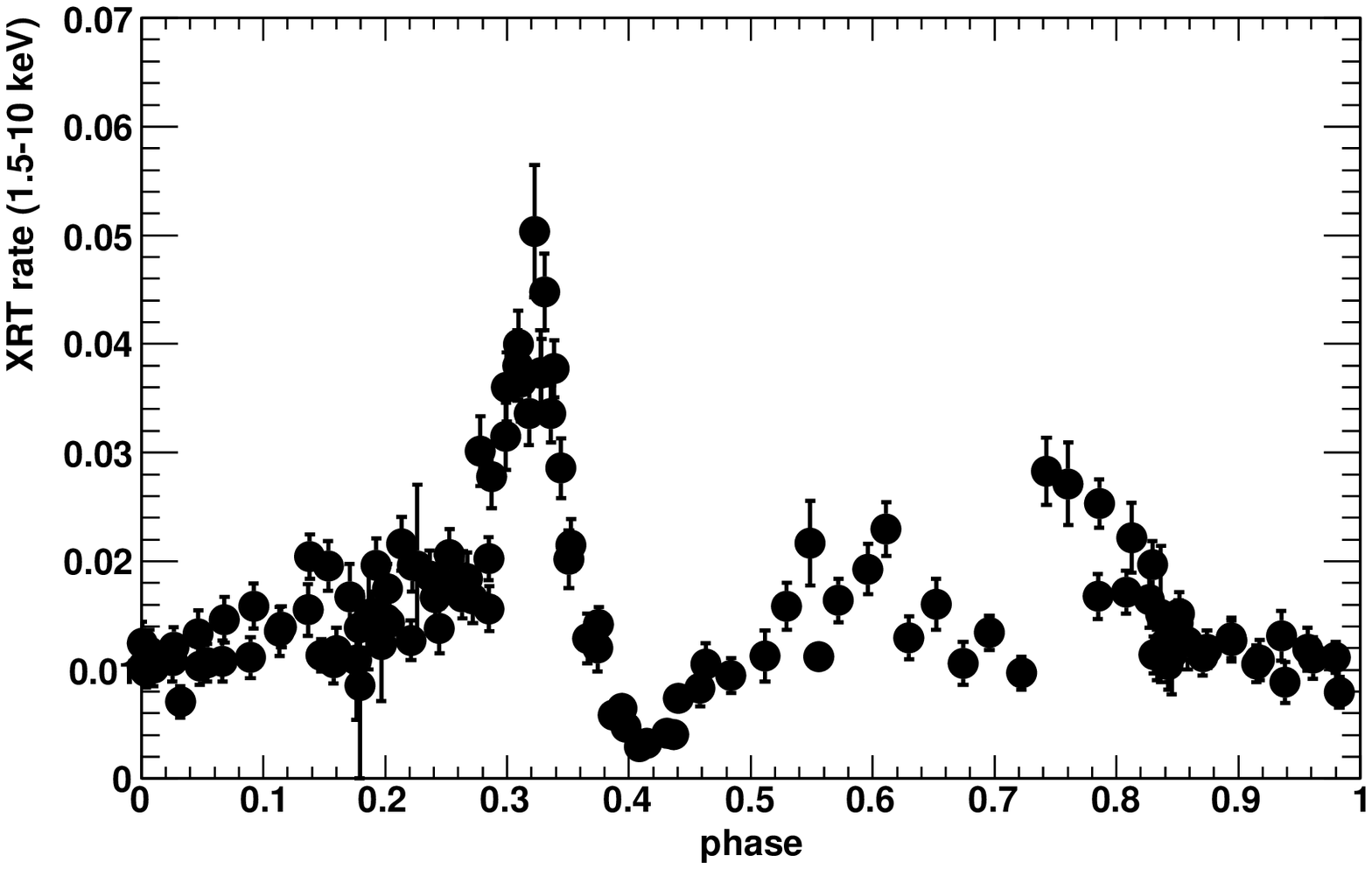}
  \includegraphics[width=5in]{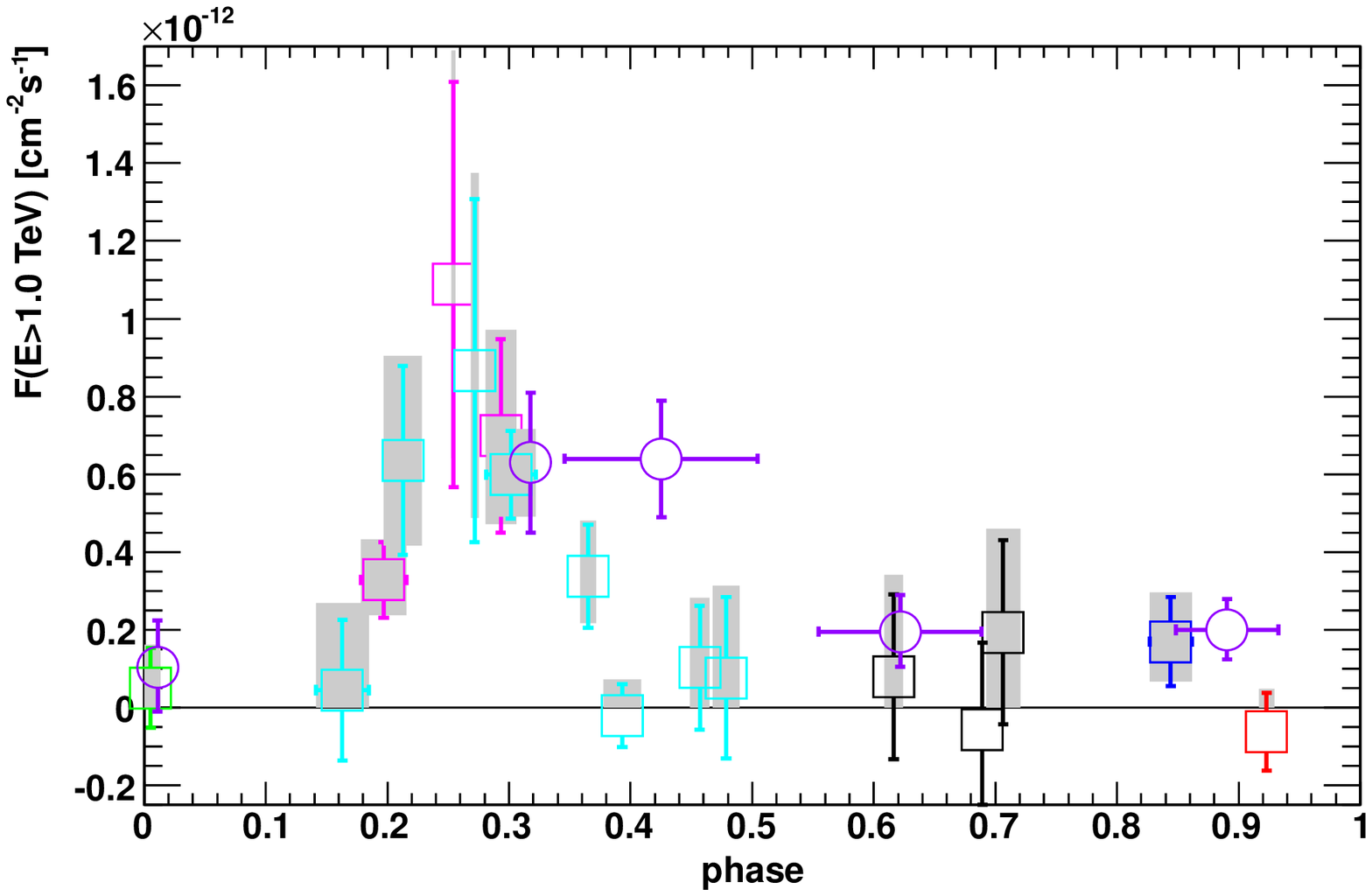}
  \caption{The Swift ({\bf Top}) and VERITAS/ H.E.S.S. ({\bf Bottom}) light curves for
  HESS~J0632+057, folded by the 321-day X-ray period \cite{maier1}. }
  \label{hessj0632}
 \end{figure*}

HESS J0632+057 was first identified as an unresolved point source of
TeV gamma rays in HESS observations of the Monoceros Loop SNR
\cite{hessj0632_detection}. The possibility that this source might be
a new TeV binary system, comprised of a massive star and a compact
companion, was noted in the original detection paper, based on the
fact that the gamma-ray emission was centered on the location of a
massive emission-line star (MWC148). Further support for this idea
came from the detection of variable radio \cite{skilton09} and X-ray
\cite{hinton09} sources at the same location, and with the discovery
of gamma-ray flux variability by VERITAS
\cite{acciari_variable0632}. 

Recently, a long-term monitoring campaign with Swift has been used to
identify a 321-day period in the X-ray emission, presumably associated
with the binary orbit \cite{bongiorno}. The X-ray emission is
characterized by a bright flare lasting for $\sim20\U{days}$, which
may be associated with a periastron passage. VERITAS and
H.E.S.S. observations are presented at this conference \cite{maier1}
which total $150\U{hours}$ over the past seven years, including a deep
VERITAS exposure around the X-ray flare in February 2011. A bright
gamma-ray flare is also observed, with the peak slightly offset from
the X-ray flare, suggesting that the gamma-ray emission fades away at
the onset of the X-ray high state. Figure~\ref{hessj0632} shows the
VERITAS/H.E.S.S. lightcurve, folded by the X-ray period.

\subsection{The Crab Pulsar}
 \begin{figure*}[!t]
  \vspace{0mm}
  \centering
  \includegraphics[width=5in]{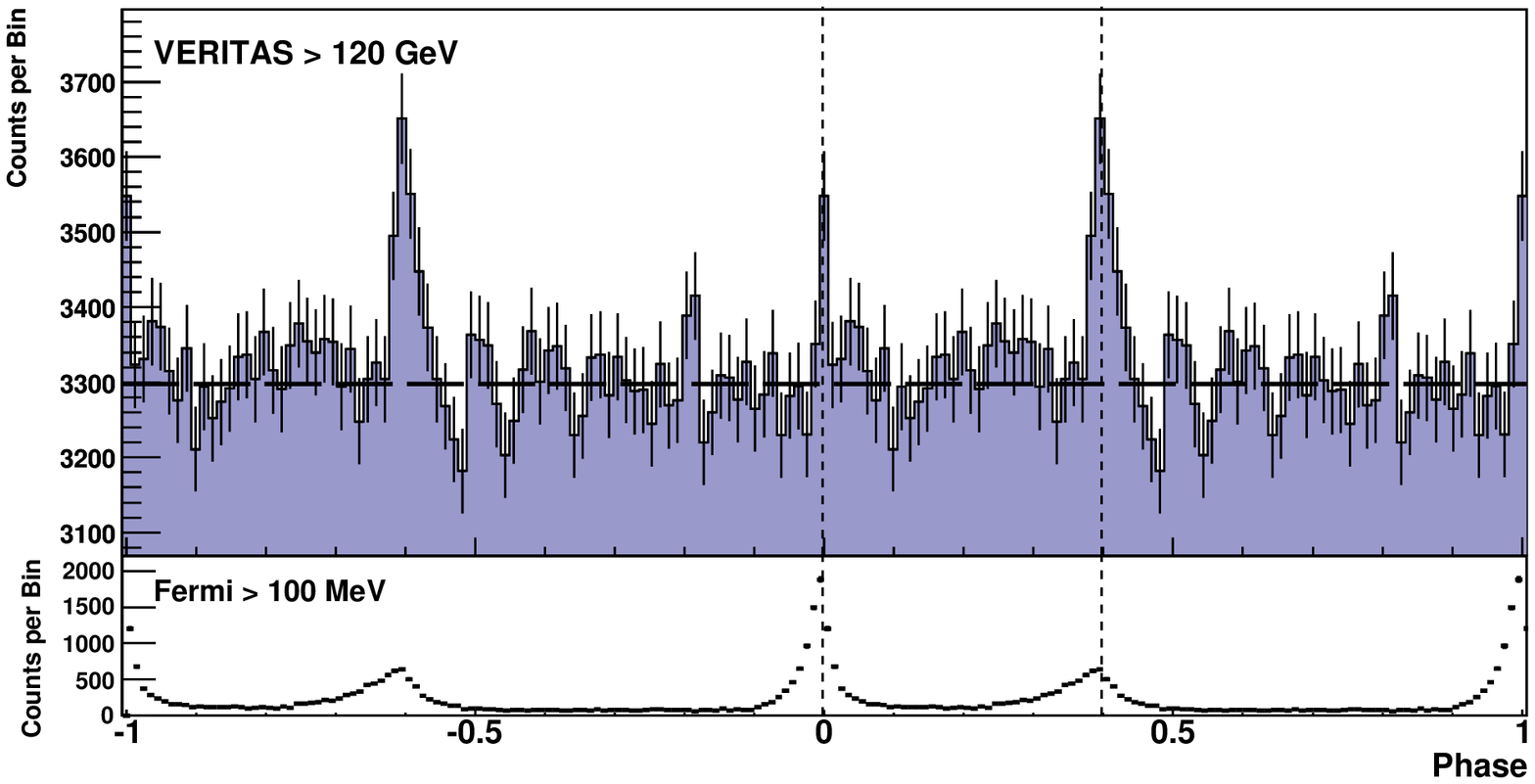}
  \includegraphics[width=5in]{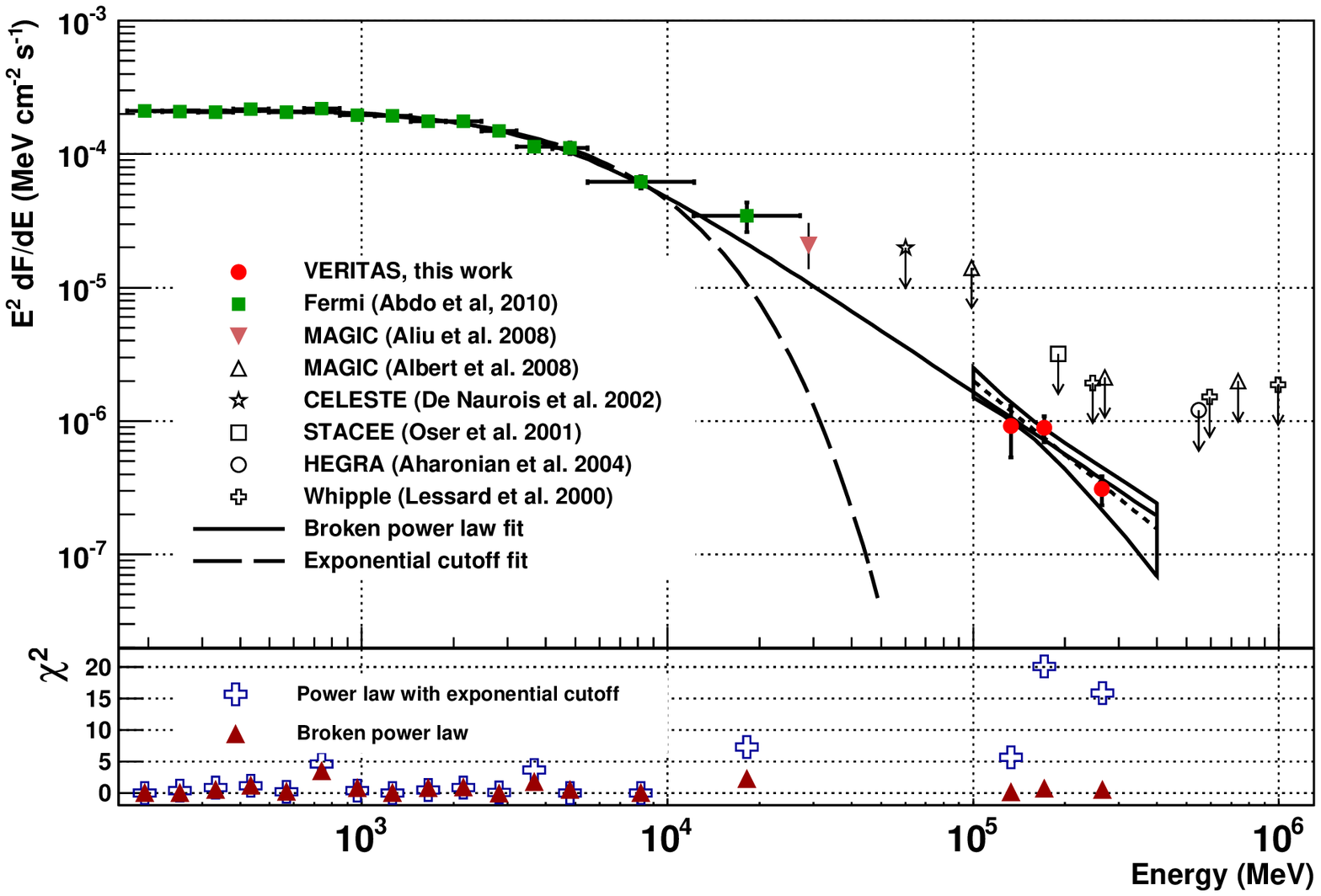}
  \caption{{\bf Top:} VERITAS measured pulse profile of the Crab
  pulsar at energies above $120\U{GeV}$ \cite{mccann,
  Science_Crab}. The pulse profile above $100\U{MeV}$ from
  \textit{Fermi}-LAT is shown underneath for comparison. {\bf Bottom:}
  Spectral energy distribution of the Crab Pulsar in gamma-rays. The
  solid line shows the result of a fit to the VERITAS and $Fermi$-LAT
  data with a broken power-law. See \cite{mccann} for details.}
  \label{crab}
 \end{figure*}

Steady emission from the Crab Nebula has provided a standard candle
for ground-based gamma-ray observatories since its detection in 1989
\cite{weekes89}. The Crab pulsar, which powers the Nebula, is among
the most energetic pulsars in our galaxy, with a spin-down power of
$4.6\times10^{38}\U{erg}\UU{s}{-1}$. Existing measurements of the
pulsar spectrum can be fit by a power-law with an exponential cut-off,
consistent with theoretical predictions based on curvature radiation
as the dominant gamma-ray production mechanism \cite{fermi_crab,
magic_crab}. 

A deep, 107 hour, VERITAS exposure of the Crab pulsar region now
reveals that the pulsar spectrum extends to much higher energies than
previously expected \cite{mccann, Science_Crab}. The emission is fit
by a power-law spectrum between 100 and $400\U{GeV}$ with an index of
$\alpha=-3.8\pm0.5_{stat}\pm0.2{sys}$. The pulse profile
(Figure~\ref{crab}) exhibits two peaks, 2-3 times narrower than those
measured at $100\U{MeV}$. The dominant pulse is observed at phase 0.4
and a smaller pulse at phase 0.1, again, in contrast to the measured
profile at $100\U{MeV}$. These results likely require a substantial
revision of our understanding of the high energy emission from
pulsars, both in terms of the location of the emission region, and the
mechanisms responsible.

\subsection{The Galactic Centre}
\begin{figure}[!h]
  \vspace{0mm}
  \centering
  \hspace{-5mm}
  \includegraphics[width=3.2in]{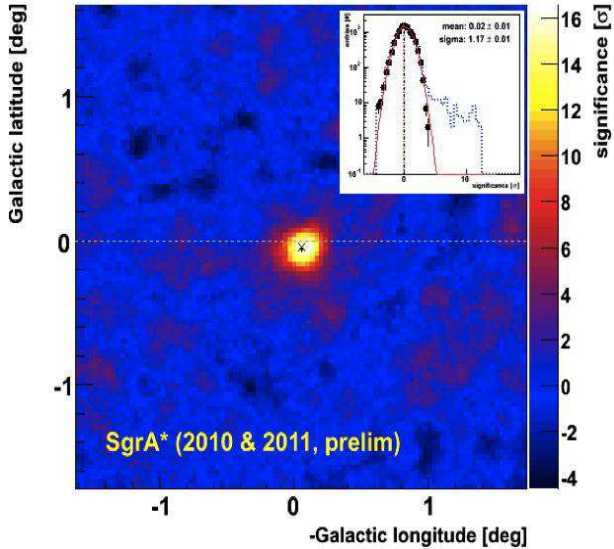}
  \caption{A significance map of the Galactic Center region, measured with VERITAS}
  \label{GC}
 \end{figure}

The Galactic Centre is a complex region at TeV energies, comprised of
a bright source in the direction of SgrA*, and extended diffuse
emission along the Galactic ridge \cite{HESS_GC, Whipple_GC,
HESS_ridge}. For VERITAS, the source culminates at an elevation angle
of below $30^{\circ}$. The path length to the Cherenkov emission from
the air shower is therefore larger than for observations at high
elevation angles, resulting in an increase in the energy threshold
above which gamma-ray initiated showers can be detected. This is
compensated by a corresponding increase in the collection area for
high energy primaries, due to the larger light pool on the
ground. Using an analysis specifically tailored to such observations,
a $25\U{hour}$ VERITAS exposure has been used to clearly detect the Galactic
Centre gamma-ray source, and to measure its spectrum above $2\U{TeV}$
\cite{beilicke} (Figure~\ref{GC}). Future observations using this method offer the
possibility of improving measurements of the high energy cut-off of
the emission spectrum.

\subsection{The Cygnus OB1 Region}

\begin{figure*}[!t]
  \vspace{0mm}
  \centering
  \includegraphics[width=5in]{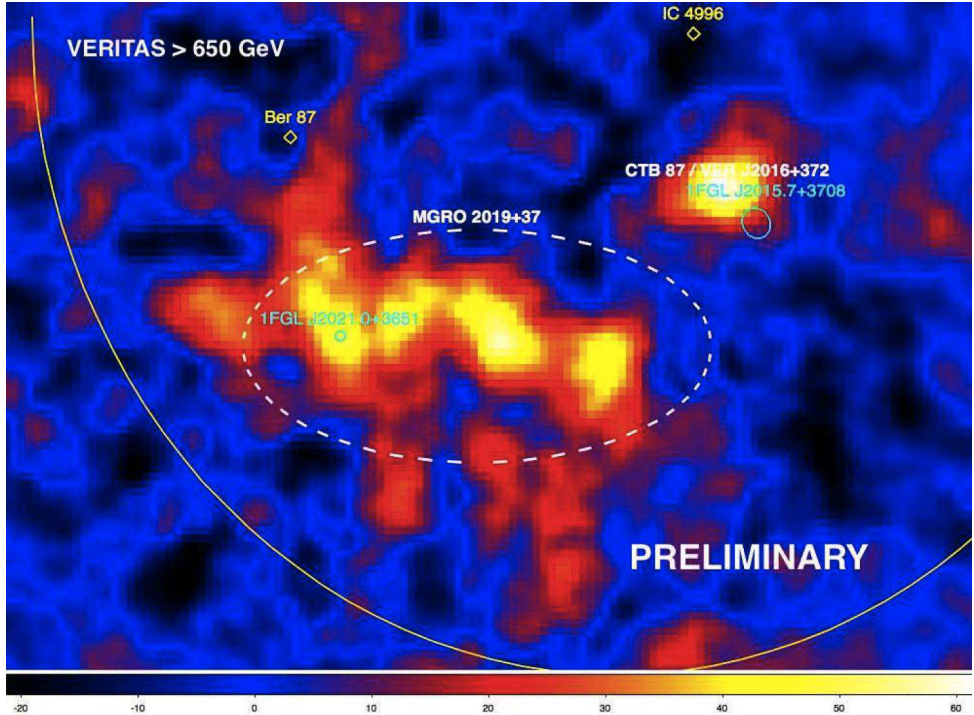}
  \caption{A map of excess gamma-ray events for VERITAS observations
  of Cygnus OB1 (indicated by the large yellow partial circle), above
  an energy of $650\U{GeV}$. See \cite{aliu} for details.}
  \label{Cisne}
 \end{figure*}

The Cygnus region hosts some of the closest and most active areas of
massive star formation and destruction in the Galaxy, providing
numerous potential sources for TeV gamma-ray production. The brightest
and most extended source in the Milagro Galactic Plane survey,
MGRO~J2019+37 overlaps with the Cyg~OB1 association, with a total flux
of $\sim80\%$ of the steady Crab Nebula flux above 12\U{TeV}, spread
over an area of $0.6^{\circ}\times1.0^{\circ}$ \cite{milagro}. At this
conference we present 75 hours of VERITAS observations centered on
this region \cite{aliu}, revealing two distinct areas of gamma-ray
excess. One of these, VER~J2016+372, is point-like, within the angular
resolution of the instrument, and consistent with the location of a
pulsar wind nebula, CTB~87, which is the likely counterpart. The other
gamma-ray excess is extended over a region which encompasses multiple
possible counterparts, including the powerful pulsar PSR~J2021+3651
and its nebula. Figure~\ref{Cisne} shows the VERITAS skymap of this
region above $650\U{GeV}$.

\subsection{Tycho's SNR}
\begin{figure}[!h]
  \vspace{0mm}
  \centering
  \includegraphics[width=3.2in]{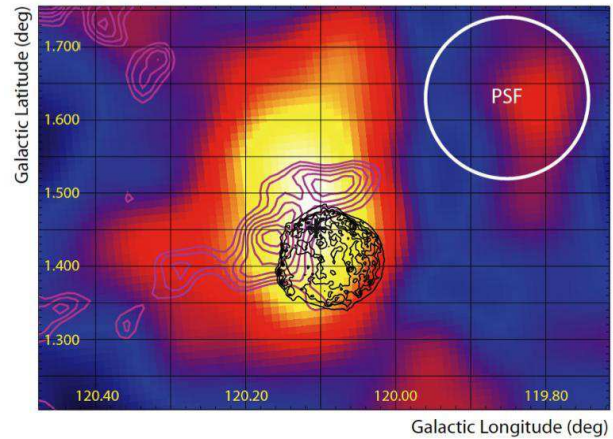}
  \caption{A map of excess gamma-ray events for VERITAS observations
  of Tycho's SNR. black contours show the Chandra X-ray emission,
  magenta contours are $^{12}$CO. The instrumental PSF is also
  indicated. See \cite{tycho} for details.}
  \label{tychomap}
 \end{figure}

Tycho's SNR is the remnant of a historical type Ia supernova, recorded
by Tycho Brahe in 1572. The SNR environment and morphology are
relatively simple, making Tycho's SNR a favored system for theorists
seeking to model particle acceleration processes in SNR blast waves
(e.g. \cite{volk08}). Evidence for nuclear particle acceleration has
already been claimed, based on detailed studies of the shock front and
contact discontinuity locations \cite{warren}. VERITAS observations of
Tycho's SNR comprise a 68\U{hour} exposure, taken between 2008 and
2010, and reveal a faint, unresolved gamma-ray source
($\sim0.9\%\U{Crab}$), significant at the level of 5.0 standard
deviations \cite{tycho}. Figure~\ref{tychomap} shows the gamma-ray
excess map, along with X-ray and $^{12}$CO emission. Dense molecular
cloud/ SNR interactions provide both the high energy particle
population and interaction target material ideal for the production of
a high energy gamma-ray signal (e.g. \cite{hewitt}). However, no direct
evidence of such an interaction exists, and the molecular emission may
simply be a chance association.

The VERITAS result alone is consistent with both leptons and hadrons
as the dominant particle population responsible for the gamma-ray
emission, although in both cases the models provide evidence for
magnetic field amplication, and hence nuclear particle
acceleration. Recent results from $Fermi$-LAT provide further
constraints and, combined with the VERITAS results, have been used to
argue strongly for a hadronic origin \cite{giordano, morlino}.

\section{The VERITAS Upgrade}
As described in the introduction, the sensitivity of VERITAS has
steadily increased since the array was commissioned. Ongoing analysis
developments promise incremental improvements in the future, but a
significant further step requires new hardware. VERITAS is currently
in the process of implementing a series of fully-funded upgrades to
the instrument \cite{kieda}. These consist of:

\begin{itemize}
\item Replacing all of the existing photomultiplier tubes in the
telescope cameras with more sensitive, super-bialkali devices.
\item Replacing the telescope-level trigger system with a higher speed, FGPA-based device. 
\item Upgrading inter-telescope networking and communications.
\item Adding instrumentation to the central pixels of each camera to
allow high-speed optical monitoring and stellar intensity
interferometry.
\end{itemize}

 \begin{figure}[!h]
  \vspace{0mm}
  \centering
  \hspace{-5mm}
  \includegraphics[width=3.5in]{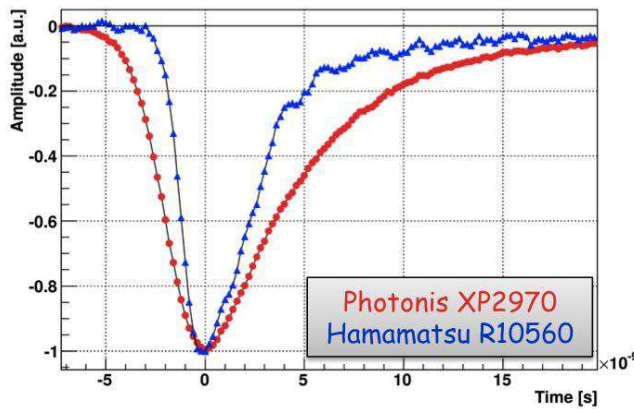}
  \caption{Measured pulse shapes for the existing (Photonis XP2970)
  and replacement (Hamamatsu R10560) photomultiplier tubes, after
  signal cable dispersion.}
  \label{pulses}
 \end{figure}

\begin{figure*}[!t]
  \vspace{0mm}
  \centering
  \includegraphics[width=5in]{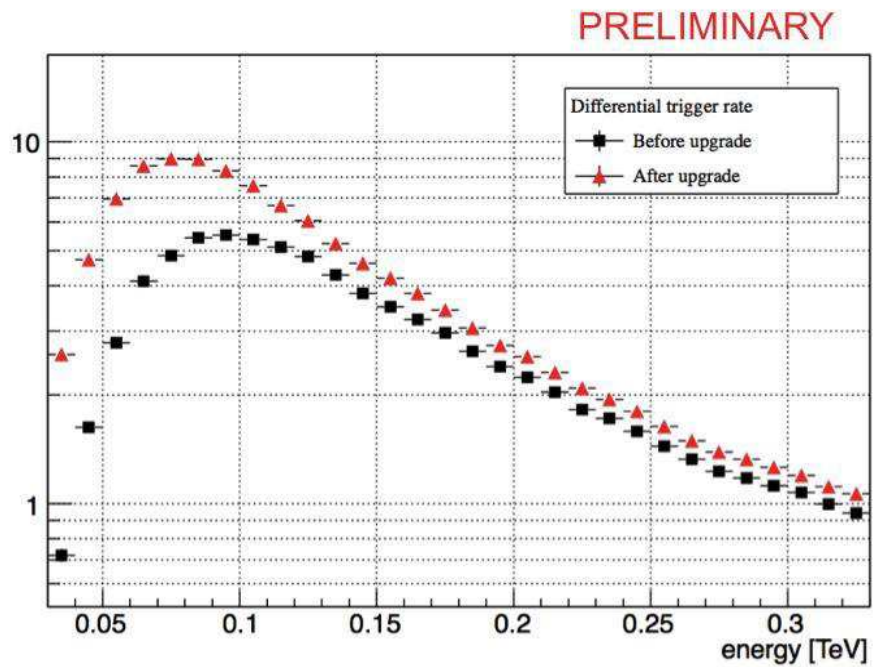}
  \caption{Monte Carlo predictions of the differential event rate for
  gamma-ray events from a Crab-Nebula-like source at the trigger level,
  both before and after the upgrade.}
  \label{trigrate}
 \end{figure*}

The first of these, the photomultiplier replacement, is expected to
have the most significant impact on the array performance. The new
photomultipliers (Hamamatsu R10560-100-20) are currently being
fabricated and tested. Initial results, both in the laboratory and in
the telescope cameras, indicate at least a 35\% improvement in
Cherenkov photon collection efficiency \cite{otte}. The Hamamatsu PMTs
are also significantly faster: Figure~\ref{pulses} shows the measured
pulse shape, in comparison with the existing Photonis
tubes. Installation of the new PMT pixels will take place during the
regular summer shutdown in 2012. All four of the upgraded trigger
systems have also been constructed and tested, and will be installed
in Fall 2011. No loss of observing time is foreseen for any of the
upgrade tasks. Figure~\ref{trigrate} illustrates the expected increase
in the differential gamma-ray event rate at the trigger level for a
source with a spectrum similar to that of the Crab Nebula. The
predicted energy threshold at the hardware level is defined as the
energy giving the peak rate in this figure, corresponding to
$75\U{GeV}$.

\section{Summary}

The results presented here constitute a brief summary of the status
the VERITAS project after $\sim4\U{years}$ of operation. Observations
are expected to continue for at least the next $\sim5\U{years}$, which
will allow us to fully exploit the VERITAS upgrade, and to complement
the Fermi-LAT effectively until the next generation of ground-based
instruments come online.\\

\section*{Acknowledgements}
This research is supported by grants from the US Department of Energy
Office of Science, the US National Science Foundation, and the
Smithsonian Institution, by NSERC in Canada, by Science Foundation
Ireland, and by STFC in the UK. We acknowledge the excellent work of
the technical support staff at the FLWO and the collaborating
institutions in the construction and operation of the instrument.


\clearpage


\begin{thebibliography}{}
\bibitem{weekes02} T. C. Weekes \etal, Astropart. Phys., 2002 {\bf 17} 221
\bibitem{holder06} J. Holder \etal, Astropart. Phys., 2006 {\bf 25}, 391
\bibitem{benbow} W. Benbow \etal, These Proceedings
\bibitem{galante1} N. Galante \etal, These Proceedings\\ (arXiv:1109.6057)
\bibitem{majumdar} P. Majumdar \etal, These Proceedings\\ (arXiv:1109.6000) 
\bibitem{errando_orr} M. Errando, M. Orr \& E. Kara, These Proceedings
\bibitem{pichel} A. Pichel \etal, These Proceedings\\ (arXiv:1110.2549)
\bibitem{weekes89} T.C. Weekes \etal, ApJ, 1989, {\bf 342},379
\bibitem{galante2}N. Galante \etal, These Proceedings\\ (arXiv:1109.6059)
\bibitem{orr} M. Orr \etal, These Proceedings
\bibitem{senturk} G. Senturk \etal, These Proceedings\\ (arXiv:1109.6035)
\bibitem{bllac_atel} R.A. Ong \etal, Astronomer's Telegram \#3459
\bibitem{mei} S. Mei \etal, ApJ, 2007, {\bf 655}, 144
\bibitem{HEGRA_M87} F.A. Aharonian \etal, A\&A, 2003, {\bf 403}, L1
\bibitem{Science_M87} V.A. Acciari \etal, Science, 2009, {\bf 325}, 444 
\bibitem{Nature_M82} V.A. Acciari \etal, Nature, 2009, {\bf 462}, 770 
\bibitem{aune} T. Aune \etal, These Proceedings
\bibitem{vivier} M. Vivier \etal, These Proceedings\\ (arXiv:1110.4358)

\bibitem{aliu} E. Aliu \etal, These Proceedings
\bibitem{maier2} G. Maier \etal, These Proceedings 
\bibitem{weinstein} A. Weinstein \etal, These Proceedings
\bibitem{hessj0632_detection} F. Aharonian \etal, 2007, A\&A, {\bf 469}, L1
\bibitem{skilton09} J. Skilton, \etal, 2009,MNRAS {\bf 399}, 317
\bibitem{hinton09} J. Hinton,\etal, 2009, ApJ, {\bf 690}, L101
\bibitem{acciari_variable0632} V.A. Acciari \etal, 2009, ApJ, {\bf 698}, L94
\bibitem{bongiorno} S.D. Bongiorno \etal, 2009, ApJ, {\bf737}, L11
\bibitem{maier1} G. Maier \etal, These Proceedings 
\bibitem{fermi_crab} A.A. Abdo, 2010, ApJ, {\bf 708} 1254
\bibitem{magic_crab} E. Aliu \etal, 2008, Science, {\bf 322} 1221
\bibitem{mccann} A. McCann \etal, These Proceedings\\ (arXiv:1110.4352)
\bibitem{Science_Crab} E. Aliu \etal, 2011, Science, {\bf 334}, 69

\bibitem{HESS_GC} F.A. Aharonian \etal, 2004, A\&A, {bf 425} L13
\bibitem{Whipple_GC} K. Kosack \etal, 2004, ApJ, {\bf 608} L97
\bibitem{HESS_ridge} F.A. Aharonian \etal, 2006, Nature, {\bf 439} 695

\bibitem{beilicke} M. Beilicke \etal, 2011, $3^{rd}$ Fermi Symposium, Rome, 2011

\bibitem{milagro}A. Abdo, A. \etal, 2007, ApJ, {\bf 658}, L33

\bibitem{volk08} H.J. V\"olk, E.G. Berezhko, L.T. Ksenofontov, 2008, A\&A, {\bf 483}, 529
\bibitem{warren} J.S. Warren, \etal, 2005, ApJ, {\bf 634}, 376
\bibitem{tycho} V.A. Acciari, \etal,  2011, ApJ, {\bf 730}, 20
\bibitem{hewitt} J.W. Hewitt, F. Yusef-Zadeh \& M. Wardle, 2009, ApJ, {\bf 706}, L270
\bibitem{giordano} F. Giordano, \etal, 2011, arXiv:1108.0265
\bibitem{morlino} G. Morlino, \etal, 2011, arXiv1105.6342

\bibitem{kieda} D. B. Kieda \etal, These Proceedings\\ (arXiv:1110.4360)
\bibitem{otte} A.N. Otte \etal, These Proceedings\\ 

\end{thebibliography}
\end{document}